\documentstyle[spie,psfig]{article} 

\title{Hydrodynamics of dense granular systems} 

\author{C. Salue\~na\supit{a,b}, S. E. Esipov\supit{a} and 
        T. P\"oschel\supit{c} 
\skiplinehalf 
\supit{a}James Franck Institute and the Department of Physics, \\
University of Chicago, 5640 S. Ellis Ave., Chicago, IL \hspace{0.5em}60637 \hspace{0.5em}USA
\skiplinehalf 
\supit{b} Departament de F\'{\i}sica Fonamental, Universitat de Barcelona,\\
Diagonal 647, Barcelona \hspace{0.5em}08028 \hspace{0.5em}Spain
\skiplinehalf 
\supit{c} Institut f\"ur Physik, Humboldt-Universit\"at zu Berlin, \\
Invalidenstra\ss e 110 
Berlin \hspace{0.5em}D-10115, \hspace{0.5em}Germany
}

\authorinfo{Further author information: \\ C.S (correspondence):
 Email: clara@hermes.ffn.ub.es; Telephone: 34-3-4021150; Fax: 34-3-4021149; 
 S.E.E: Email: 
 esipov@franck.uchicago.edu;  
 T.P: Email:thorsten@sp10.physik.hu-berlin.de; Telephone:49-(0)30-2093-7896;
 Fax: 49-(0)30-2093-7638. 
 }

\begin{document} 
  \maketitle 

%%%%%%%%%%%%%%%%%%%%%%%%%%%%%%%%%%%%%%%%%%%%%%%%%%%%%%%%%%%%% 
\begin{abstract}
The properties of dense granular systems are analyzed from a hydrodynamical 
point of view, based on conservation 
laws for the particle number density and linear momentum. We discuss
averaging problems associated with the nature of such systems and the
peculiarities of the sources of noise. We perform a quantitative study
by combining analytical methods and numerical results obtained by 
ensemble-averaging of data on creep during compaction and molecular 
dynamics simulations of convective flow. We show that numerical integration
of the hydrodynamic equations gives the expected evolution for the time-dependent
fields.
\end{abstract}

%>>>> Please include a list of keywords right after the abstract 


%%%%%%%%%%%%%%%%%%%%%%%%%%%%%%%%%%%%%%%%%%%%%%%%%%%%%%%%%%%%%
\section{INTRODUCTION} 

The interest in granular materials goes back to the early sixties,
when industries needing to process powders required more and more control
over the quality of their products. One may imagine a variety of shapes for
mixers and  many  different ways to drive the grains inside into motion; a 
down-to-earth approach would consist of an answer for the question: in which 
of these mixers should I spend money? This can be done if one can compare
their performances by means of realistic models that reproduce
true flow patterns. 

In order to answer aspects of this problem, and based on the gas-like
appearance of the compounding particles, attempts were made to apply
a hydrodynamical description to granular systems.
It has been more than a decade since the pioneering contributions to the
understanding of granular materials from a hydrodynamical point of 
view\cite{Haff83,Haff86}. This understanding, however, is still far from 
being complete. We
will address here several questions of theoretical and practical interest,
namely fluctuations and averaging problems intrinsic to granular nature, 
modelling of dense granular flows by means of hydrodynamic equations
and boundary conditions, comparisons with experimental data and results
from molecular dynamics simulations, and fluctuations and mixing problems.  
%%%%%%%%%%%%%%%%%%%%%%%%%%%%%%%%%%%%%%%%%%%%%%%%%%%%%%%%%%%%%
\section{THE SOURCES OF FLUCTUATIONS} 

The behavior of a granular mass is fundamentally different from
that of typical fluids.  For instance, consider ordinary hydrodynamic
fluctuations: the lack of an intermediate length scale \---much larger
than the typical diameter of the grains but much smaller than the size
of the system, makes the thermodynamic
limit unnatainable and therefore hydrodynamic fluctuations may subsist
in the continuum limit. In addition, the fact that solid grains in a dense 
arrangement can't be regarded as 
points in any length scale leads to a second source of fluctuations; it operates
at distances much larger than the typical diameter of the particles and is
related to the appearance of extensive arrangements inside
the system. Given a mean density not far away from the close-packing limit,
there is a fraction of the free volume which is still available and can
be distributed in many ways. Some of these configurations will flow, some
will not allow net motion, so
different actual realizations in the configuration space of the particles
may lead to departing dynamic properties and, ultimately, in generating
completely different time-sequences. This contribution is what we call
non-local noise. Both, local and non-local noise, are always present
but their relative importance strongly depends on the forcing applied to the
system, measured by the parameter ${\cal F}g = f/\rho g$, $f$ being the volume density of
the forcing and $g$ the acceleration of gravity.

Therefore, one can distinguish two sources of fluctuations and two 
types of averaging, one over local noise and the other over different
configurations or realizations. 

We propose the existence of two regimes; in the weak-forcing limit, ${\cal F}g 
 \stackrel < \sim  1$, the non-local component of the noise dominates, and consequently
one expects very long  relaxation rates and non-self-averaging quantities.
In this limit, only ensemble averaging is meaningful, as in every possible
rea\-li\-za\-tion the system explores a small region of the configurational space.
The glass-like behavior is most apparent. In the opposite limit, ${\cal F}g \gg
1$, the system is not easily trapped in an immobile arrangement, and one
can safely assume that in a sufficiently long time it explores the 
representative part of the 
configuration space. Time ave\-ra\-ging can substitute ensemble averaging only in this
limit. Consistently with this picture, the cri\-ti\-cal density, $\rho_c$, is not
unique in general, but is a distributed quantity depending on the 
configurational state, $\Gamma$. Actually, we shall see that  expe\-ri\-mental data
on compaction at weak forcing display quenched behavior, 
and the final density may vary over more than 10\%.
Instead, our numerical study of granular convection under strong forcing indicates
a much narrower histogram of maximal achieved densities, $\rho_c$, despite the
fact that the number of particles and the used number of samples for averaging
were much smaller. Consider that for high forcing, the mean density evolves
within a comparatively wide margin and the system very rarely, if ever, visits 
any of the corresponding trapping configurations. In the mean-field limit, 
one can consider  $\rho_c$ as an unique constant, which would correspond to the
strict close-packing limit.

In spite of the complexity of this picture,  we show that it is possible to
 understand both weak and strong forcing limits within the frame of 
hydrodynamic equations, which are in general stochastic equations including
both local and non-local noise. It is beyond of the scope of this paper
a detailed analysis of the properties of such equations, which is extensively
done elsewhere. Our aim is to present some results of our study of the
evolution of the mean hydrodynamic fields, comparing them with their observed 
behavior in a sample of cases.

%>>>> \cite{} provides references as superscripts
%>>>> \citenum{} provides references as numbers in the normal font 

%%%%%%%%%%%%%%%%%%%%%%%%%%%%%%%%%%%%%%%%%%%%%%%%%%%%%%%%%%%%%
  \section{HYDRODYNAMIC EQUATIONS} 
 
%%-----------------------------------------------------------
By hydrodynamic equations we mean balance equations for mean mesoscopic
hydrodynamic fields. We will focus on the equations for the conservation
of mass and linear momentum, leaving the energy balance for a later discussion.
As said in the previous section, the lack of an intermediate length scale  
\---contrary to what happens in simple
fluids for ins\-tan\-ce, containing a sufficiently large number of particles
such that local fluctuations fade away, adds on the above exposed problem of 
non-local fluctuations due to the intrinsic granular nature of the system, 
operating at length scales where hydrodynamic fields can already be defined.
The former can be modelled as the usual additive stochastic contribution to the
dissipative flows and comes related to the existence of some kind of 
"temperature", the latter enters via distributed kinetic coefficients 
(depending on the configurational state $\Gamma$). In the continuum
approach, conservation of mass and linear momentum read, in the Stokes 
approximation,
        \begin{eqnarray}
        \label{mass}
        \partial_t\rho + \nabla \cdot \rho {\bf v} &=& 0,\\
        \label{linmom}
        \rho\partial_t v_i &=& 
        \frac{\partial\sigma_{ij}}{\partial{x}_j} + {f}_i,
        \end{eqnarray}
where ${f}_i$ are the components of the volume density of forcing.
Note that these equations express very basic laws. However, they are
valid in the usual form when the averages decouple, that is, when
the average of the "microscopic" linear momentum equals the product
of the averages of $\rho$ and ${\bf v}$. However, due to the mesoscopic
nature of the averaging procedure, this is not guaranteed. 
It will work if the local fluctuations are small enough. We shall discuss
the fact that local fluctuations of the velocity field are very small
in dense granular flows, except the very few moments when the
granular mass undergoes dramatic changes, and close to the boundaries of
the system. 
 
As for the terms constituting the stress tensor $\sigma_{ij}$ a few comments
are in order. Provided that the granular particles may be modelled as 
sufficiently hard spheres, we neglect any elastic contribution  other than 
that introduced by pressure effects, and assume that the non-equilibrium
part of $\sigma_{ij}$ is a local functional of the derivatives of the local 
velocity,
        \label{stress}
        \begin{equation}
        \sigma_{ij} = - p\delta_{ij} + \eta(\rho,\Gamma)
        \Big(\frac{\partial{v}_i}{\partial
        {x_j}} + \frac{\partial{v_j}}{\partial{x_i}} - \frac{2}{3}\delta_{ij}
        \nabla\cdot{\bf v}\Big)
         + \zeta(\rho,\Gamma)\delta_{ij}\nabla\cdot{\bf v} + \xi_{ij}. 
        \end{equation}
where $p$ is the pressure and $\eta$ and $\zeta$ the shear and volume
viscosities, respectively.

%%-----------------------------------------------------------
\subsection{The role of temperature and the pressure term} 

Our molecular dynamic simulations indicate that the evolution of the 
velocity field in dense granular flow is nearly independent from the granular 
temperature, defined as 
the mean fluctuational part of the velocity. Effectively, well inside the bulk,
the quantity $\delta v/ v$, measuring such fluctuational deviations from
the mean velocity $v$, is typically about six orders of magnitude smaller
than close to the boundary. This and other evidences allow us to suppose that
granular flows in dense systems can't be sustained by a temperature-based
mechanism alone \---unlike Rayleigh-B\'{e}nard convection in simple fluids, for instance. Note
that the equation for the energy balance has been omitted; consistently
with the observation that the granular temperature plays no significant
role, its evolution appears decoupled from the previous system of equations\footnote{ 
Actually, we observed that the rate of energy release, coming from inellastic 
collisions, was clearly correlated with the viscous heating (see next
subsection for the model for the viscosity). These terms
are therefore responsible for the evolution of the temperature field, since
the energy loss is proportional to some power of the granular temperature, $T$.
We checked the functional dependency, resulting in the expected $T^{3/2}$.}.  
Similarly, one cannot account for elastic contributions in
the high density limit only by using a thermal pressure. Therefore, one
has to model such terms by means of an artificial equation of state which must
help to resolve the delicate limit $\rho \rightarrow \rho_c$.
We assumed the most simple dependence, $p = p_0/(1-\rho/\rho_c)$, where
$p_0$ represents a certain constant,
in our numerical integration of the hydrodynamic equations in the strong
forcing regime.
%%----------------------------------------------------------- 
\subsection{The viscosity}.

Accordingly, the model that we adopt for the viscosity is not thermal,
but glass-like. In dense clusters, in order to move, a complex rearrangement 
of particles has to occur making use of voids. Similar properties are exhibited
by glasses. Available experimental data and our numerical results indicate the 
presence of a factor $\exp[c/(1-\rho/\rho_c))]$ in the mean flow rates, 
where $c$ is a dimensionless number. This formula is related to 
the V\"ogel-Fultcher law for glasses\cite{Vogel-Fultcher}; it measures 
the number of attempts needed 
for one step in the direction of average flow in a dense gra\-nu\-lar system.
Sufficiently close to $\rho_c$, we then expect a shear viscosity of the form
\begin{equation}
\label{visc}
\eta(\rho, \Gamma)  = \eta_0(\rho, \Gamma) 
   \exp\Big(\frac{c}{1 - \rho/\rho_c(\Gamma)}\Big).
\end{equation}
and a similar dependence for the bulk viscosity, $\zeta(\rho,\Gamma)$.
%%%%%%%%%%%%%%%%%%%%%%%%%%%%%%%%%%%%%%%%%%%%%%%%%%%%%%%%%%%%%%%%%%%%%%
  \section{TEST OF THE HYDRODYNAMIC EQUATIONS} 

We discuss some representative examples  
of each regime. For the weak forcing limit, available experimental
data on compaction of sand during tapping experiments\cite{Knight95}
provide the necessary reference.
For the strong forcing limit we perform extensive ensemble averaging of
samples generated by molecular dynamic (MD) simulations of vertical and
horizontal shaking,
comparing the resulting hydrodynamic fields with those generated integrating 
numerically the hydrodynamic equations. We shall also
show  analytical results on the cycle-averaged velocity profiles
in vertical shaking that fit experimental profiles\cite{NMR}.   

%%-------------------------------------------------------------------
  \subsection{Weak forcing limit. Application to compaction experiments.}
  
\begin{figure}[htb]
        \centerline{\psfig{file=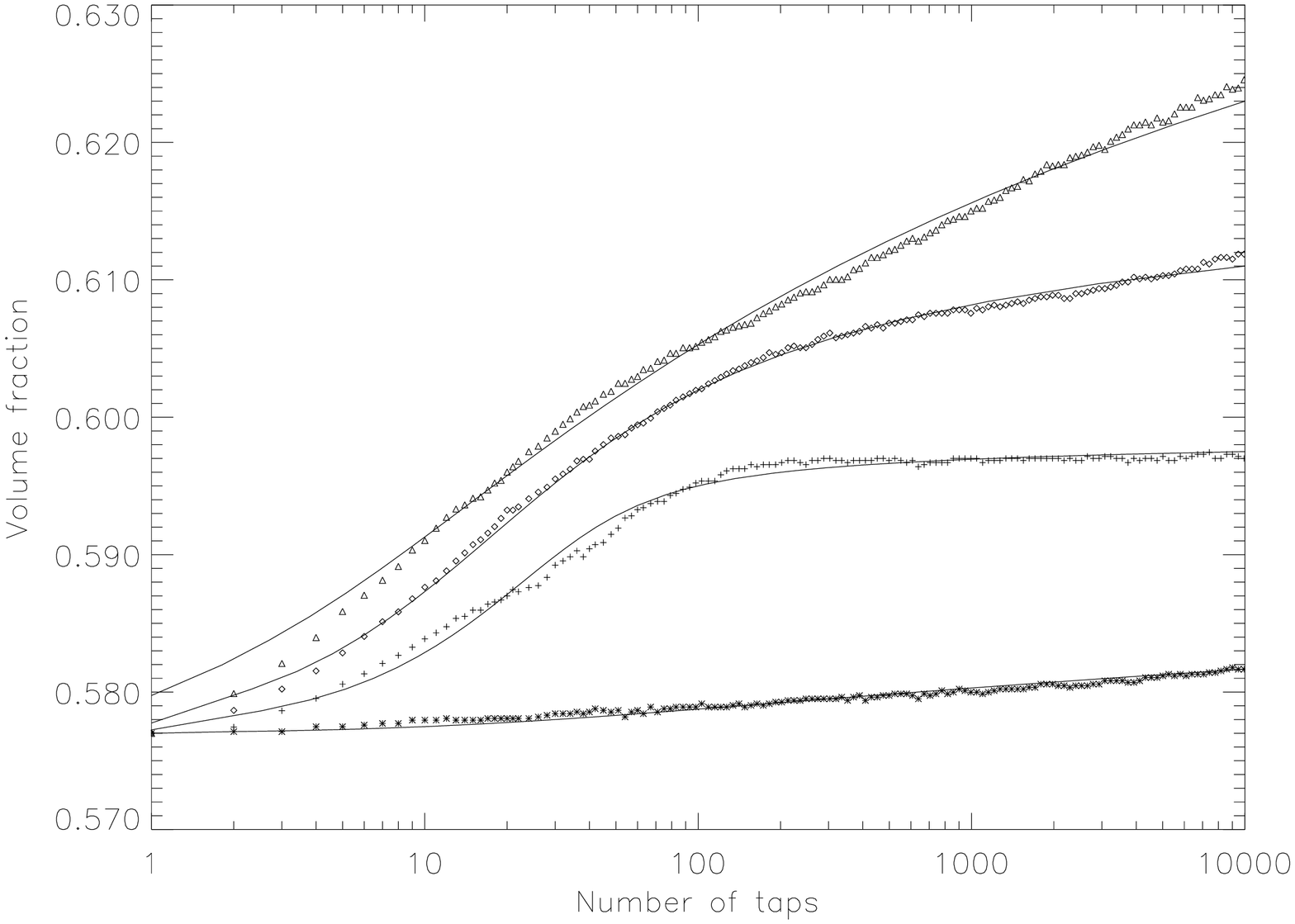,width=9cm,angle=90} 
                    \psfig{file=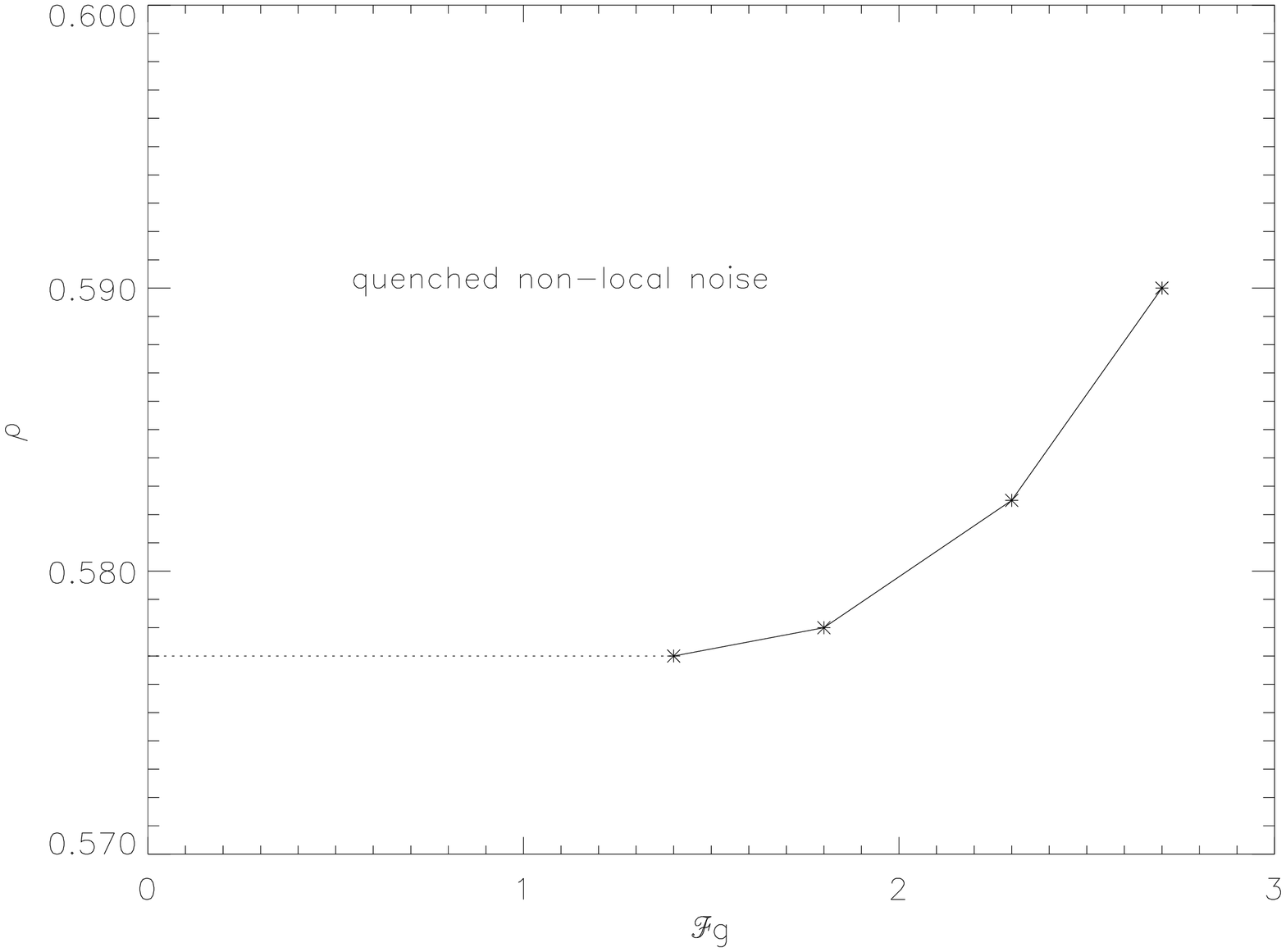,width=9cm,angle=90}}
        \caption[capt1] 
%>>>> use \label inside caption to get Fig. number with \ref{}
        { \label{fig:fits}      
  a) Fits of the experimental curves of the density dependence on
  the number of taps, from Ref.~\citenum{Knight95}.
  ${\cal F}g = 1.4, 1.8, 2.3, 2.7$
  from the bottom to top.  The following parameter
  values were used to fit the curves in the same order:
  $\langle{F}\rangle/\eta_0 =
  (6.9 \cdot10^{-8}, 4.1 \cdot10^{2}, 3.1 \cdot10^{1}, 1.1 \cdot10^{-2})$,
  $c = (0.92, 0.029, 0.18, 1.35)$, $\rho_c = (0.5985, 0.599, 0.62, 0.67)$. \\
  b) Region of quenched non-local noise as a function of the density and
  the forcing parameter {\cal F}g.}     
        \end{figure}

These results provide additional support for the existence of non-local
noise, and evidence that the mean flows are of hydrodynamic nature even
in very dense limits. Beginning with a loosely packed sand at volume fraction 
$\rho_0 = 0.57$ the authors report a logarithmic density growth, 
        \begin{equation}
        \label{compaction}
        \rho_c - \rho(t) = \frac{A}{1 + B\ln(1+t/\tau)},
        \end{equation} 
where $A, B, \tau, \rho_c$ are four fitting parameters. It can be shown
that it is possible to retrieve such a dependence by integration of hydrodynamic
equations. Omitting further details about calculations and average over non-local
noise, one finds after integration of the 1-dimensional version of (\ref{mass},\ref{linmom})
at late times
        \begin{equation}
        \label{compth}
        \bar{\rho_c} - \rho(t) = \frac{c}{\ln[(\int_0^tdtF)/c\bar{\eta}_0]},
        \end{equation}
$\bar\rho_c$ and $\bar{c}$, already averaged over non-local noise, 
are constants which depend on, say, 
the amplitude of forcing, but do not change over time. The quantity
$F$ is related to the integral of the density of forcing and is left 
unspecified. Equation (\ref{compth}) can be compared with experimental fit,
(see Fig.~\ref{fig:fits}a) 
assuming $\int_0^tdtF = t\langle F\rangle$, where the average is taken over
a period of repeated tapping.
We find $A/B = \bar{c}$, $\tau = c\bar{\eta}_0/\langle F\rangle$.
This is a  three parameter fit. It demonstrates that 
hydrodynamics may be used for analyzing experimental
data.  Different fitting values of $\bar\rho_c$, $\bar{c}$ support the 
assumption that granular configurations with different $\rho_c$, $c$ 
(dif\-fe\-rent states $\Gamma$)
do not communicate at weak forcing.
Now we can be more specific in what we mean by "weak" forcing.
Note that the fit is satisfactory at late times. At
early times that situation is different: the higher is the
value of ${\cal F}g$ the longer it takes before the fit is
any good. If we plot the density when the deviation from
the fit is a few percent versus the forcing parameter we get 
Fig.~\ref{fig:fits}b. 
It suggests that the fits are good when the assumption
of quenched values of $\rho_c$ and $c$ is fulfilled. Therefore,
one may expect that the non-local noise is quenched above a
certain line in the $(\rho, {\cal F}g)$ plane.

%-------------
%-------------    
%%-------------------------------------------------------------------
  \subsection{Strong forcing limit. Results from MD simulations.}

This case is object of a more complete study. Consider for example
periodic vertical shaking of sand under gravity.  
It is well known that sand in such conditions develops
ty\-pi\-cal convective rolls, with particles going upwards inside, and downwards
along the vertical walls. The motion is evidenced, for example, by the
bulging colored stripes  resulting from Magnetic Resonance Imaging experiments\cite{NMR}, 
which represent cycle-averaged displacements. Again, it is possible
to show that the system of equations (\ref{mass},\ref{linmom}) can be integrated
to give a good fit of the experimental data. We will omit here the
detailed derivation, which can be found in Ref.~\citenum{ESP96}. 
Imposing the geometry of a tall container of width {\cal L} in which the density slightly 
decreases with height, assuming Eq.~(\ref{visc}) for the 
viscosity and the incompressibility condition (see step 4, below,
for the reasons of such approximation), one finds 
\begin{equation}
\label{expt}
v_z(x,z) = v_0 e^{k^2zl_z}\left[1 - 
kL\frac{\cosh(kx)}{\sinh(kL)}\right].
\end{equation}
which is solution of Eq.~(\ref{linmom}) for the cycle-averaged vertical
velocity, in a two dimensional geometry. Here $l_z$ represents the scale
of variation of the viscosity in the linear approximation, $k$ an
arbitrary inverse length scale and $v_0$ a characteristic amplitude of the
velocity, unspecified. See Ref.~\citenum{NMR} for experimental evidence of
Eq.~(\ref{expt}).

A major understanding of the motion requires, however, a time-resolved analysis, 
and we show how this can be done via molecular dynamics simulations.
We will not discuss details about the simulations here, it suffices to say
that we used a polydisperse sample of 2000 soft spheres in a rectangular
container, and the chosen values for the friction parameters reproduce
correctly the experimental MRI images mentioned above. ${\cal F}g$ was about 2 in
the study of vertical shaking, and 9 in the case of horizontal shaking.
We choose the first for presentation, although the results 
can be extended to the horizontal shaking, leaving aside some
peculiarities which are not worth to comment here. 
The following steps summarize the procedure we used, which is schematized
in the chart of Fig.~\ref{fig:chart}:
%-------------
        \begin{figure}[htc]
        \centerline{\psfig{file=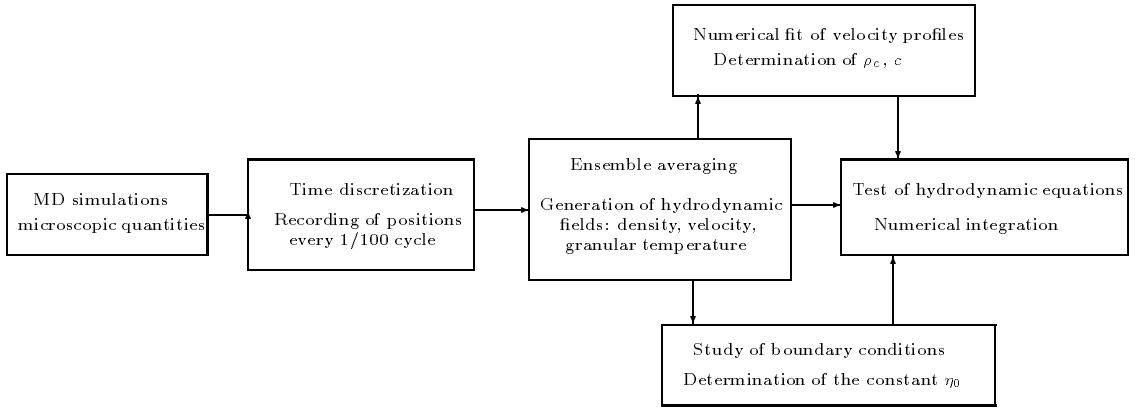,width=18cm,angle=0}} 
        \caption[capt0]
        { \label{fig:chart} 
        Scheme of the procedure followed for the test of the hydrodynamic
        equations.}      
        \end{figure} 
%-------------

\begin{figure}[htc]
        \centerline{\psfig{file=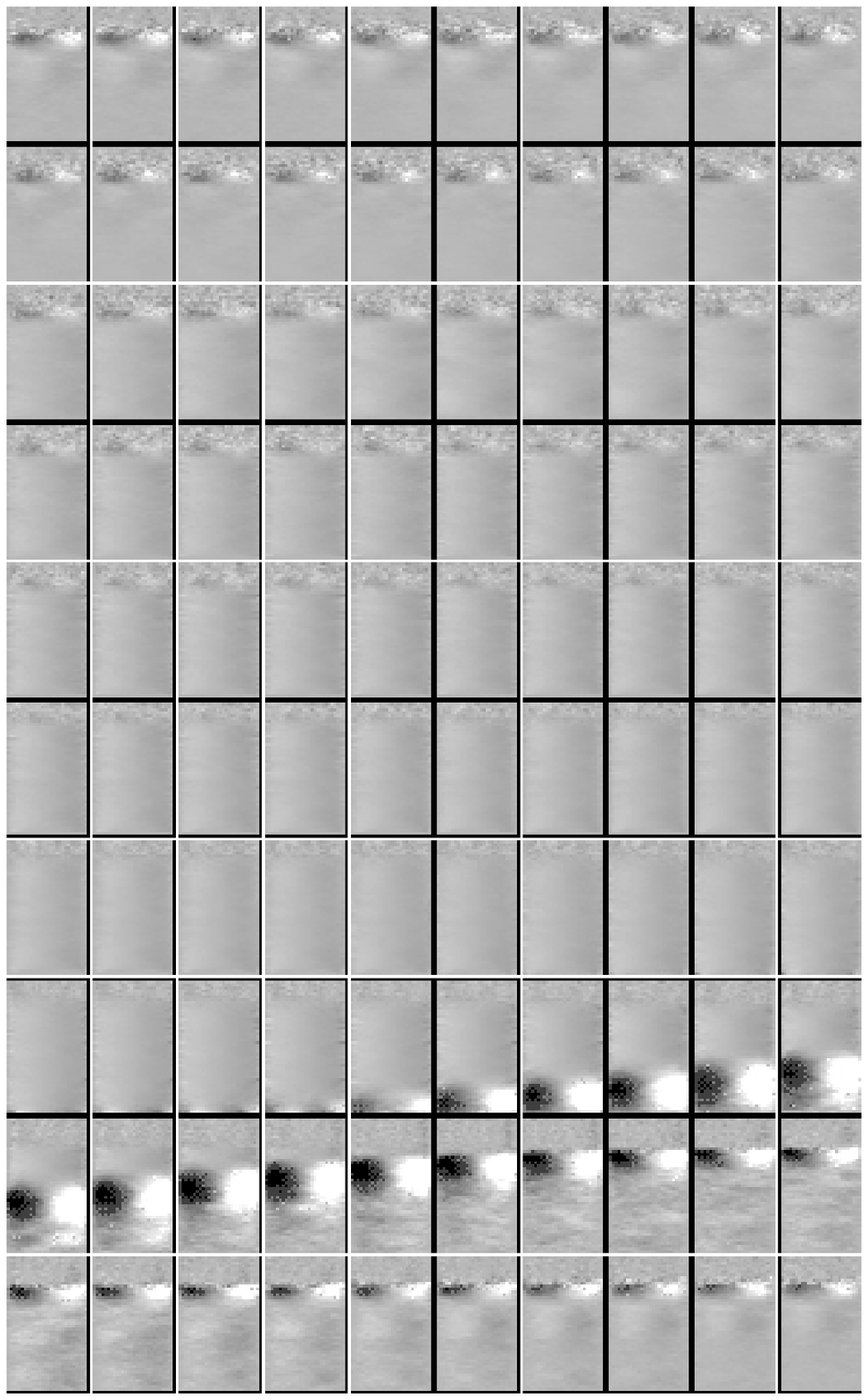,width=7cm,angle=0} 
                    \psfig{file=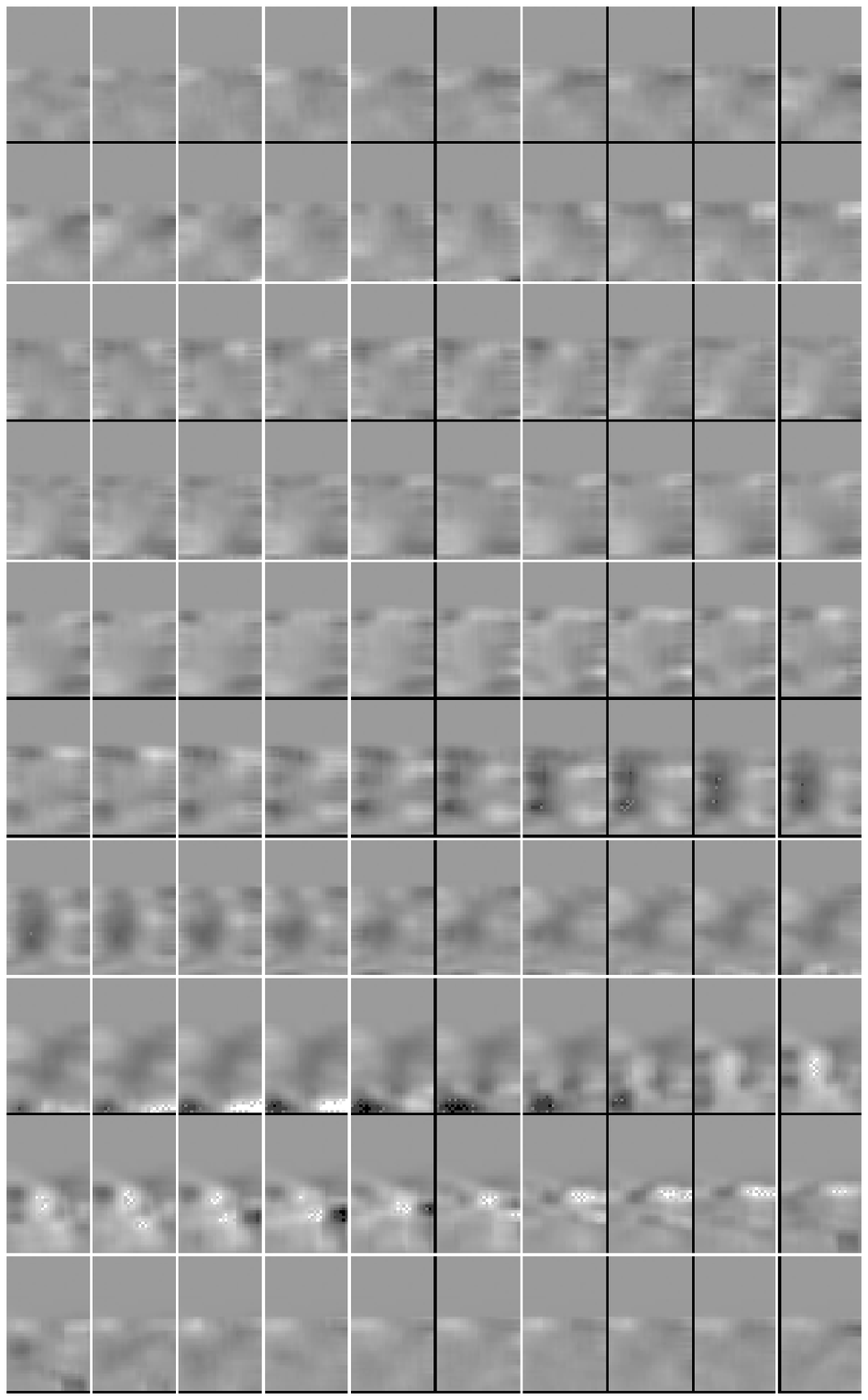,width=7cm,angle=0}}
        \caption[capt2]
        { \label{fig:vx} a) (left)
  Ensemble-averaged horizontal velocity, $v_x$ 
  of granular material 
  in the course of vertical shaking. $10\times10$ frames taken during
  one cycle of oscillation are positioned 
  from left to right and from top to the bottom. The frames of highest
  contrast indicate the collision of the granular system with the bottom
  wall, with compression of the material propagating upwards. b) (right)
  Evolution obtained by 
  numerical integration of the
  mean-field hydrodynamical equations (\ref{mass},\ref{linmom}) with the
  boundary conditions obtained in step 5. In both pictures dark areas
  represent negative values and white ones positive values of the velocity.}     
        \end{figure}

\begin{figure}[htc]
        \centerline{\psfig{file=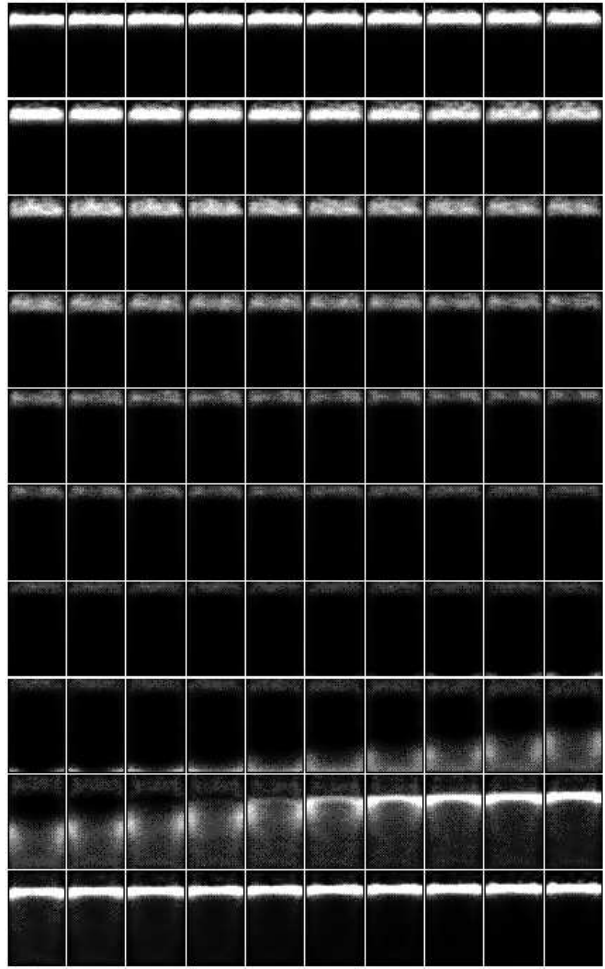,width=8cm,angle=0}} 
        \caption[capt3]
        { \label{fig:temp} 
         $10 \times 10$ sequence for the granular temperature 
         (ensemble-averaged fluctuational velocity) in one cycle of
         horizontal shaking. The positioning of frames is the same as in 
         Fig.~\ref{fig:vx}. White areas indicate high temperature regions.
         By comparing with Fig.~\ref{fig:vx}, observe that the granular mass 
         moves in the opposite direction of the temperature gradient. }  
        \end{figure}

\begin{figure}[htc]
        \centerline{\psfig{file=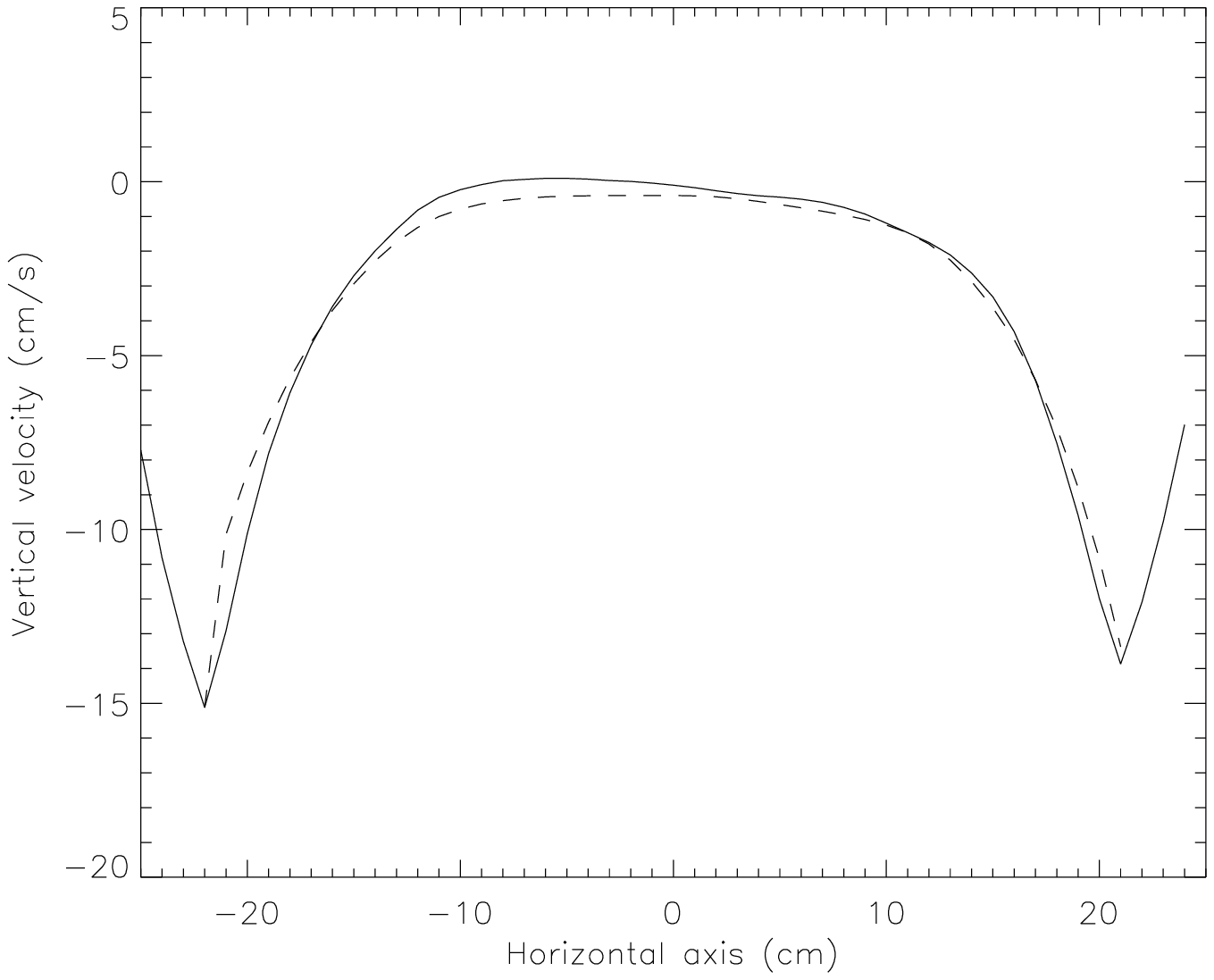,width=8cm,angle=0} 
                    \psfig{file=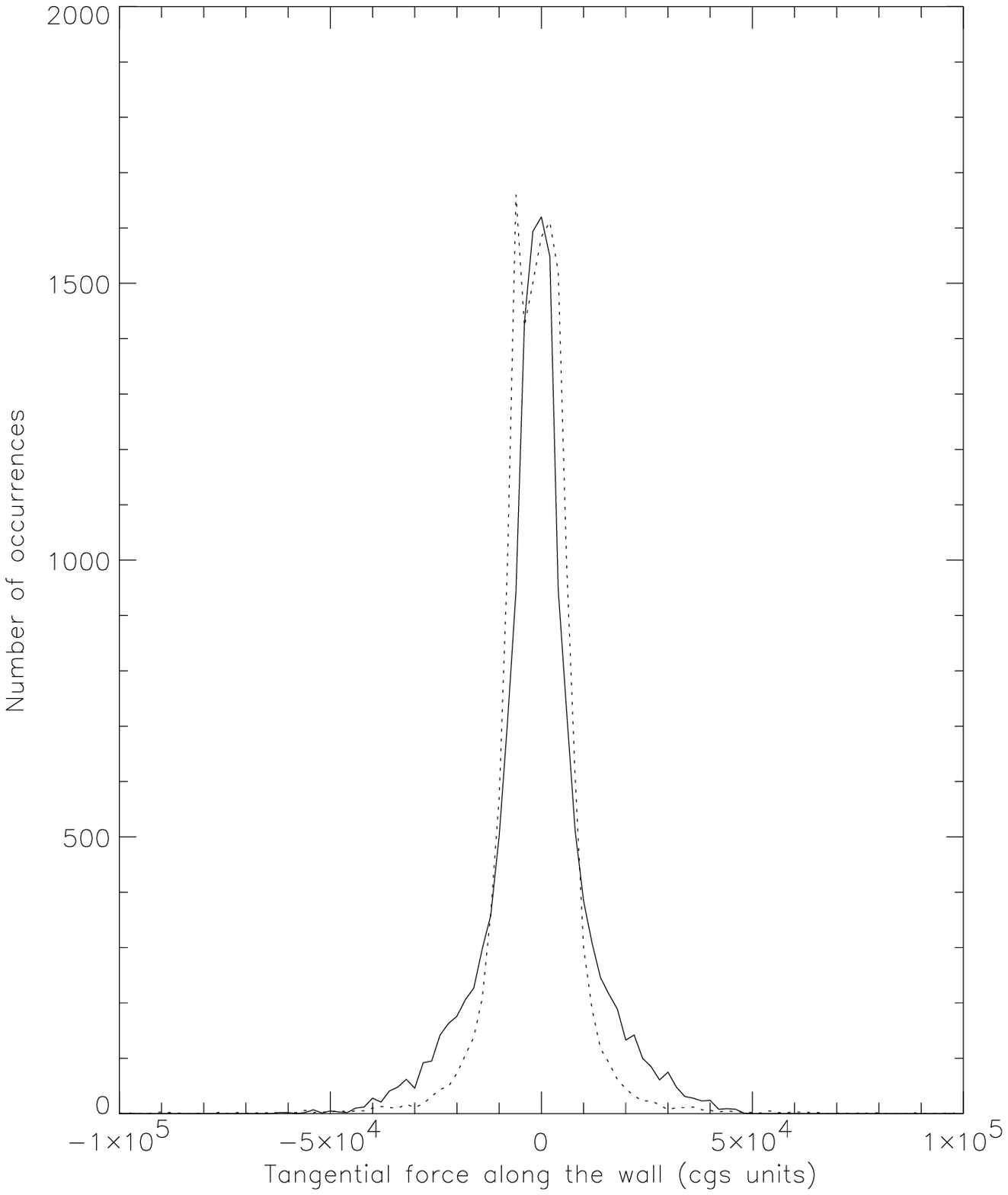,width=7cm,angle=0}}
        \caption[capt4]
        { \label{fig:visco} a) (left) Fit of the ensemble-averaged profiles of
  vertical velocity, $v_y(x)$, in frames 55-70 (numbered in the same way
  as in Fig.~\ref{fig:vx}). In the fitting procedure the
  values for  the constants $c$=0.15 and $\rho_c$= 1.01 
  max$(\rho)$ were found. Dramatic changes at the
  boundaries are due to density discontinuities accompanying
  downward sliding of sand along the side walls. b) (right) Comparative histograms of the
  distribution of tangential force density
  acting at a wall (solid line) and the non-diagonal component
  of the hydrodynamic stress tensor, $\sigma_{\tau{n}} =
  \eta(\rho)(\partial{v}_{\tau}/\partial{n})$, where $n$ and $\tau$
  stand for normal and tangential directions, respectively.}     
        \end{figure}

\begin{figure}[htc]
        \centerline{\psfig{file=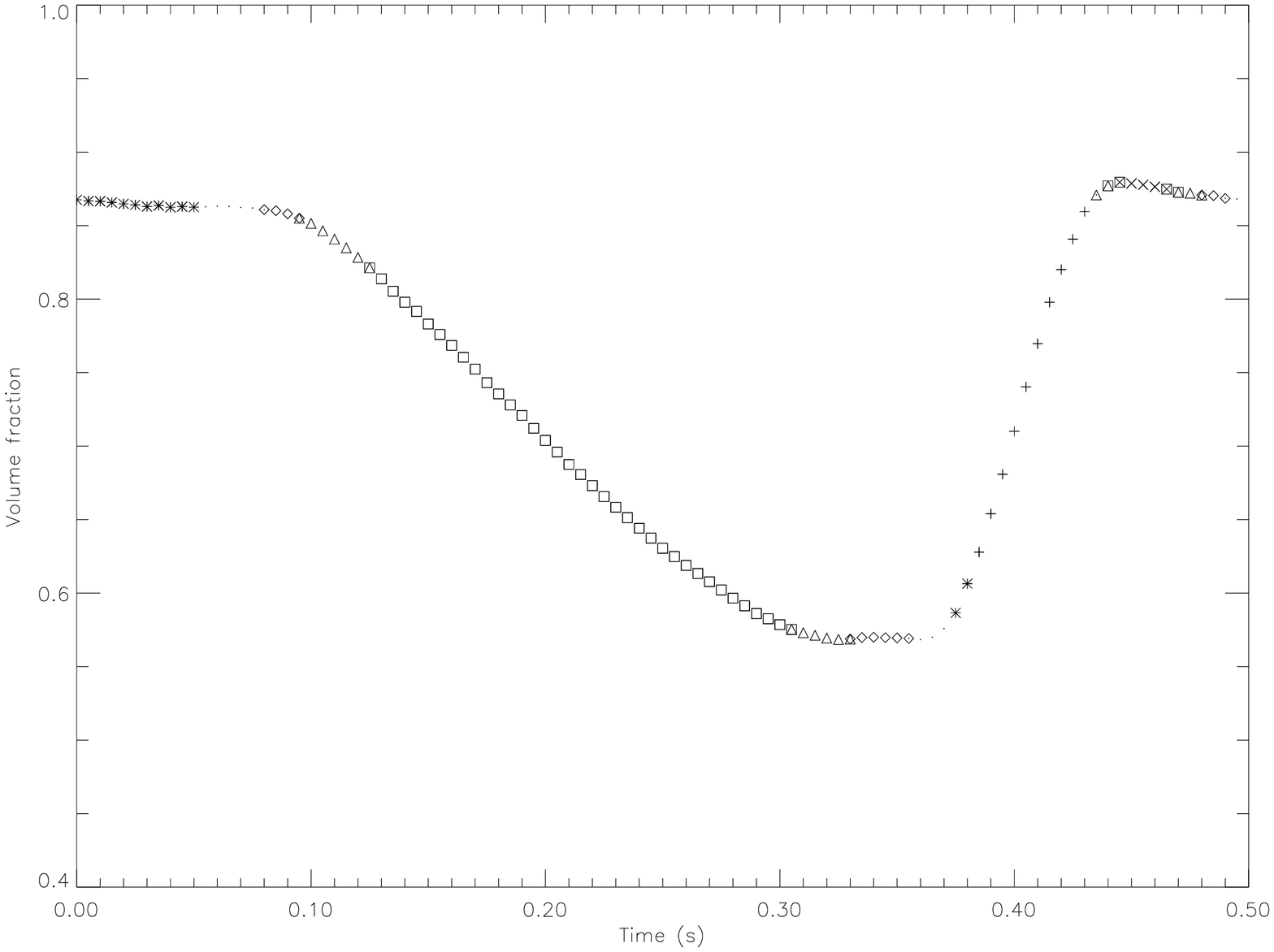,width=12cm,height=7cm,angle=90}} 
        \centerline{\psfig{file=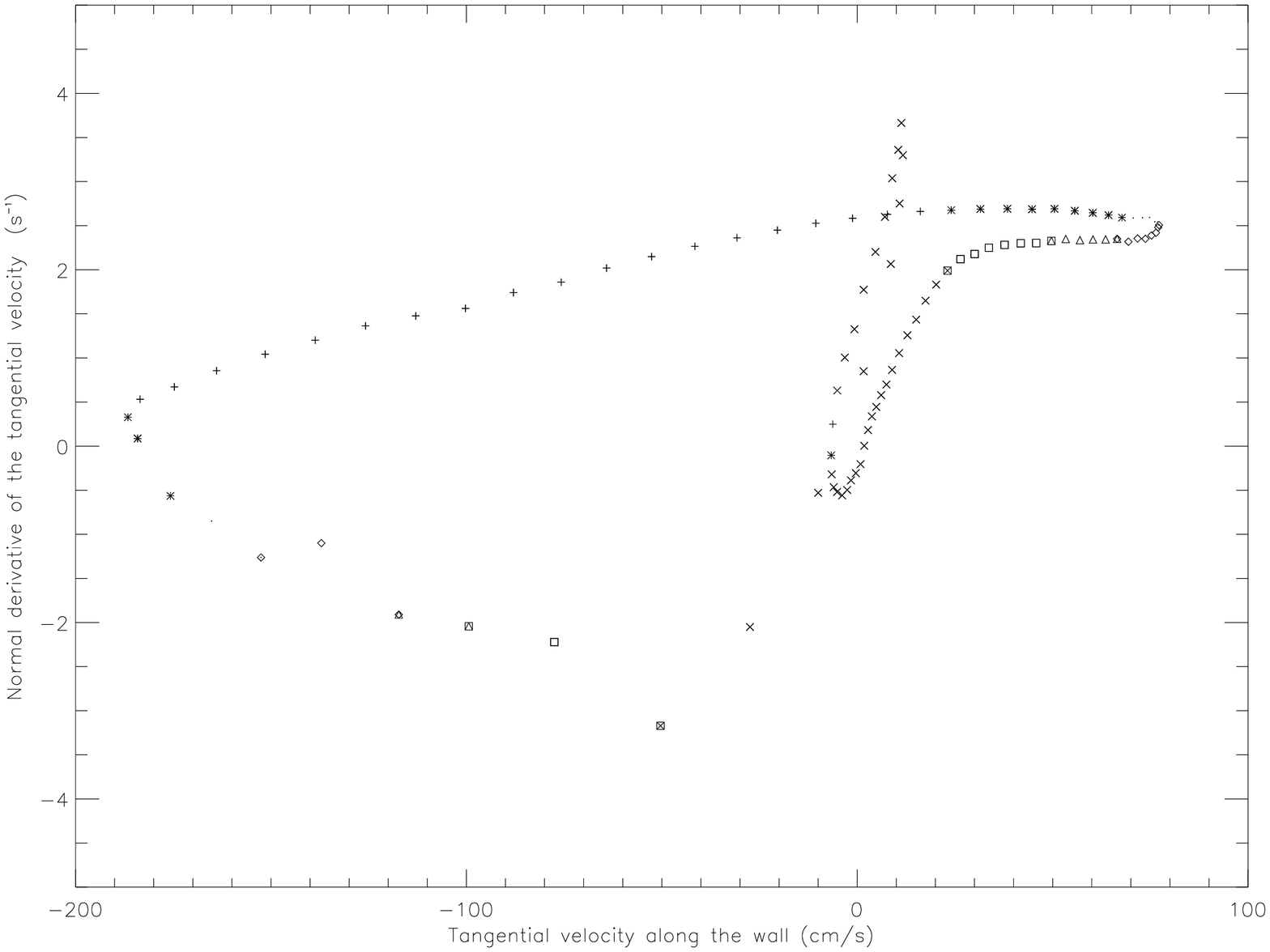,width=12cm,height=7cm,angle=90}}
        \caption[capt5]
        { \label{fig:bc} a) (above) Averaged dependence of granular 
  density close to the vertical walls during one
  period of vertical shaking. The decrease in boundary 
  density starts with the take-off of the granular material from the
  bottom of the container and reaches its maximum just before its
  landing (maximum acceleration downwards). Different
  symbols indicate the value of the off-diagonal component of the
  stress tensor, $\sigma_{\tau{n}}$. There are seven symbols used
  in linear proportion to the increasing value of  $\sigma_{\tau{n}}$:
  (+, *, $\cdot$, $\Diamond$, $\triangle$, $\Box$, $\times$). 
  b) (below) The same data plotted in a different representation
  ($\partial v_\tau/\partial n$ vs $v_\tau$).
  Plotting symbols now account for the value of granular density,
  in the same ascending order in the density
  range $0.55 \le \rho \le 0.9$.}        
        \end{figure}

\begin{enumerate}
\item Time discretization. Each period of shaking was divided in 
an equal number of frames, where positions of particles were recorded.   

\item Spatial discretization and averaging. Using a high resolution
grid, the container was divided in cells of the size of the order of one 
particle. Time averaging \---which can replace ensemble averaging in this case,
was performed with data of the cor\-res\-pon\-ding frames of more of 100 periods
of shaking.

\item Mean density, velocity and temperature (mean fluctuational
velocity) were obtained and displayed. Visualization was done by means 
of IDL movies, showing smooth, well behaved fields. The sequences revealed 
unsuspected details about the motion that the granular mass experiences 
during a cycle, totally hidden when a cycle-average is performed (which
leads to the typical convective rolls and has confused the sand community
for long time).  
  As an example, we reproduce in 
Fig.~3a the horizontal component of the velocity, $v_x$. Observe that the
motion is  complex and unexpected, in the sense that one cannot infer
from the sequence of pictures of $v_x$ (neither from $v_y$, not shown)
the direction of the global motion. The evolution of the granular 
temperature is shown in Fig.~\ref{fig:temp}. Since the darkest regions represent
the lowest values, it is obvious that its importance is restricted
to the region close to the free surface, and in the
bulk, to a limited number of frames. 

\item Test of the hydrodynamic equations. By using the mean hydrodynamic
fields obtained in 3., the system of equations (\ref{mass},\ref{linmom}) 
was checked.
Selected frames provided fitting values for $c$ and $\rho_c$  
(see Fig.~\ref{fig:visco}a).
$\eta_0$ was found to be about 300 cpoise by comparing histograms of the
tangential force and the velocity gradient close to the walls 
(Fig.~\ref{fig:visco}b), 
whereas the observation that the flow was mostly divergence-free allowed 
us to neglect the effects of $\zeta$.

\item Study of boundary conditions. We obtained effective boundary 
conditions for the flow  that reproduce to some extent the assumptions
of microscopic friction during collisions, but we also found that the
motion of sand along the vertical walls comes accompanied by dramatic
periodic changes in the density and the stress (Fig.~\ref{fig:bc}).

\item The previous results were used to integrate numerically
the system of hydrodynamic equations. In Fig.~\ref{fig:vx}b we show comparatively
the sequence obtained for $v_x$. Observe that, apart from a qualitative
agreement (including the order of magnitude of the velocity fields)
there is room for improvement.
\end{enumerate}      
%-------------
%-------------
%-------------
%-------------
%-------------
%-------------
%-------------
%-------------
%%%%%%%%%%%%%%%%%%%%%%%%%%%%%%%%%%%%%%%%%%%%%%%%%%%%%%%%%%%%%%%%%%%%%%
  \section{FLUCTUATIONS AND MIXING}

\begin{figure}[htc]
        \centerline{\psfig{file=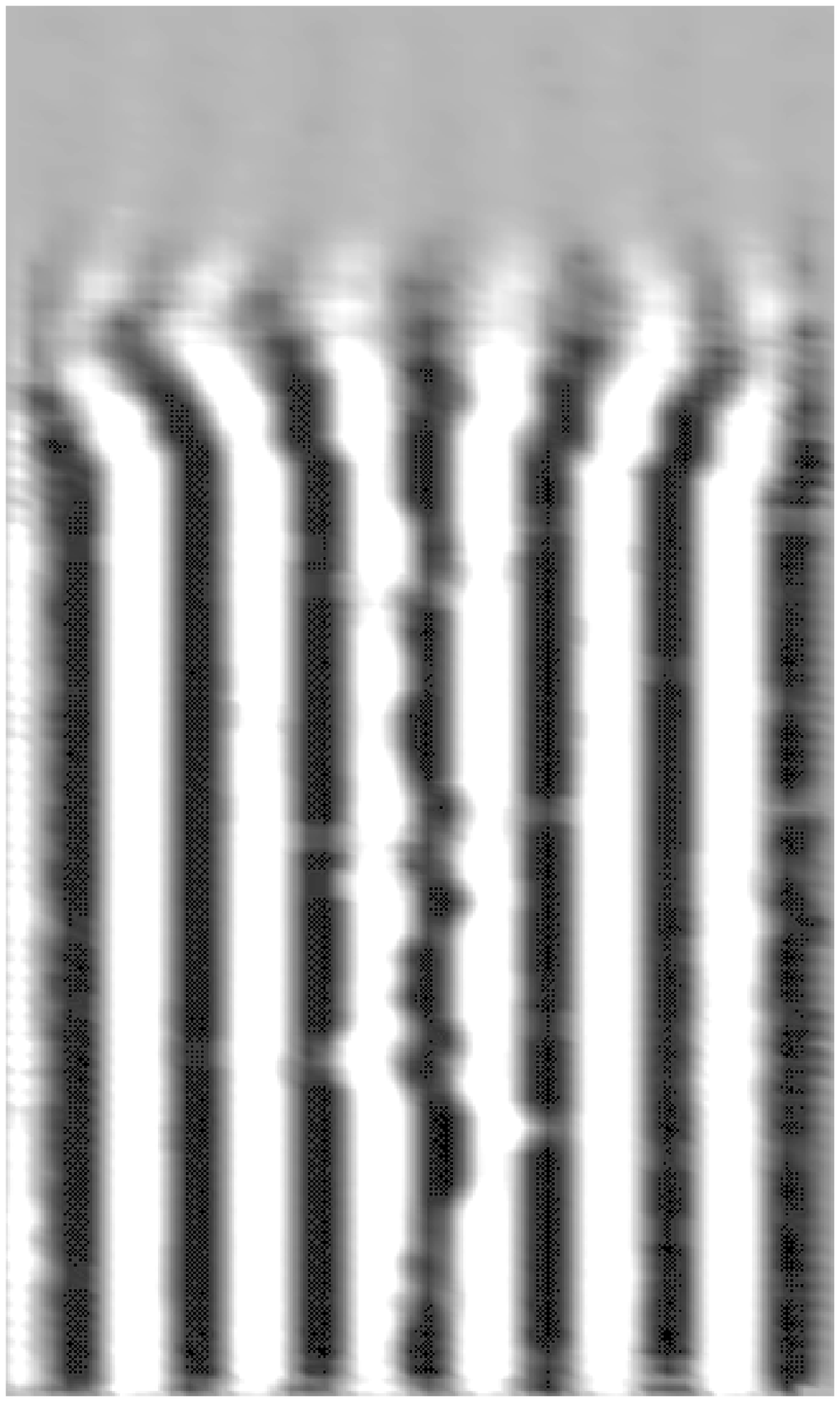,width=5cm,angle=0} 
                    \psfig{file=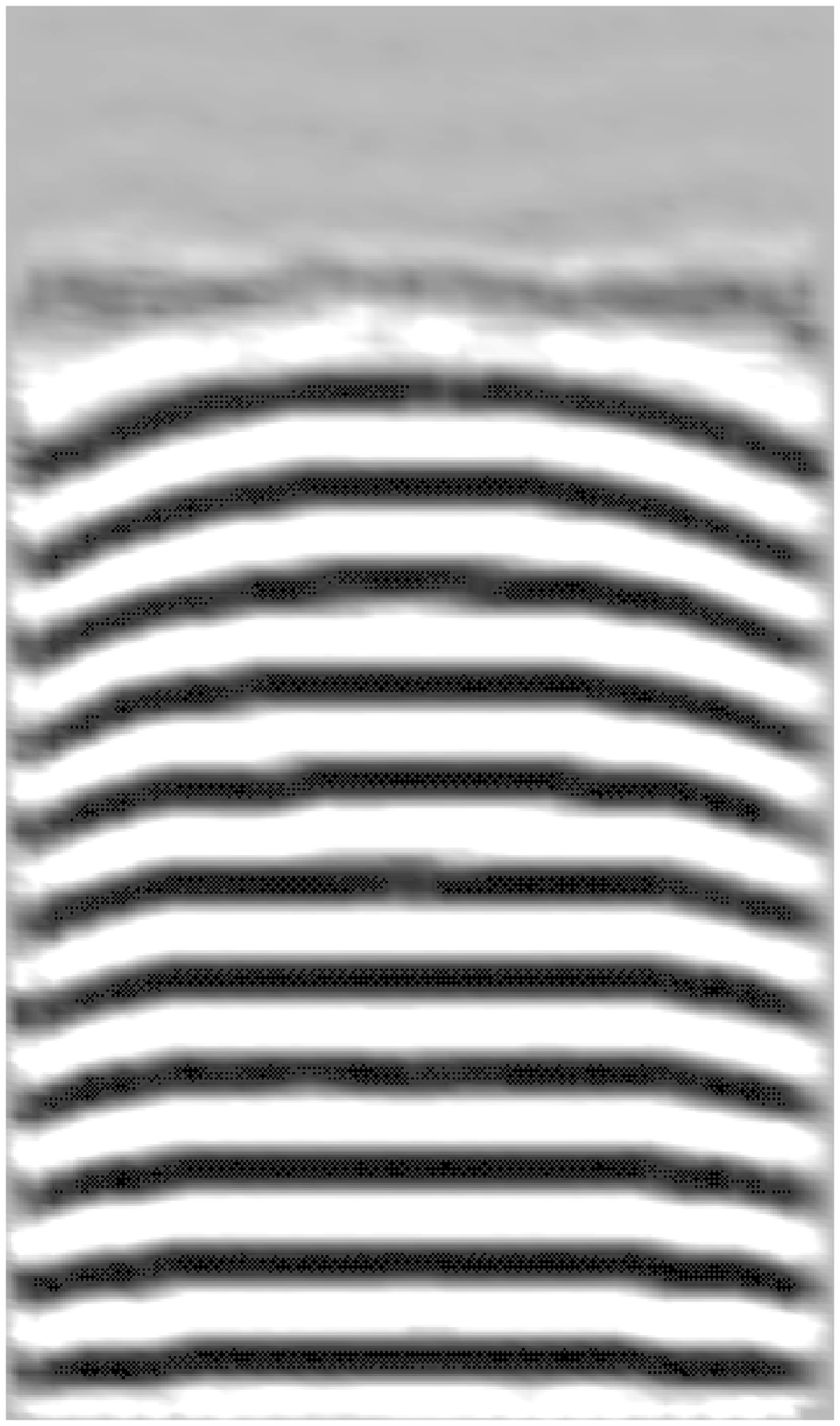,width=5cm,angle=0}} 
                    
                    \vspace{1cm}
         \centerline{\psfig{file=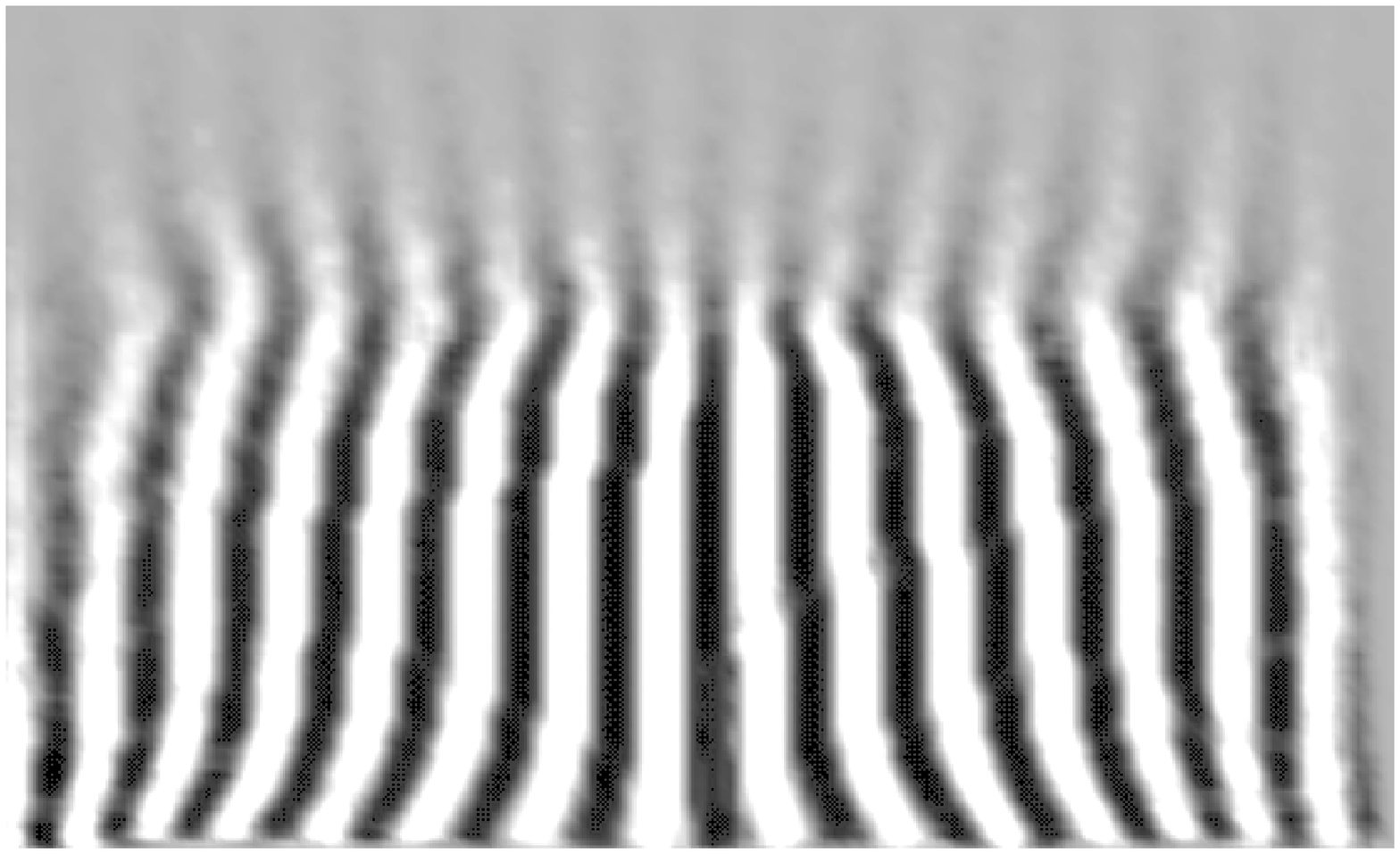,width=8cm,angle=0}
                    \psfig{file=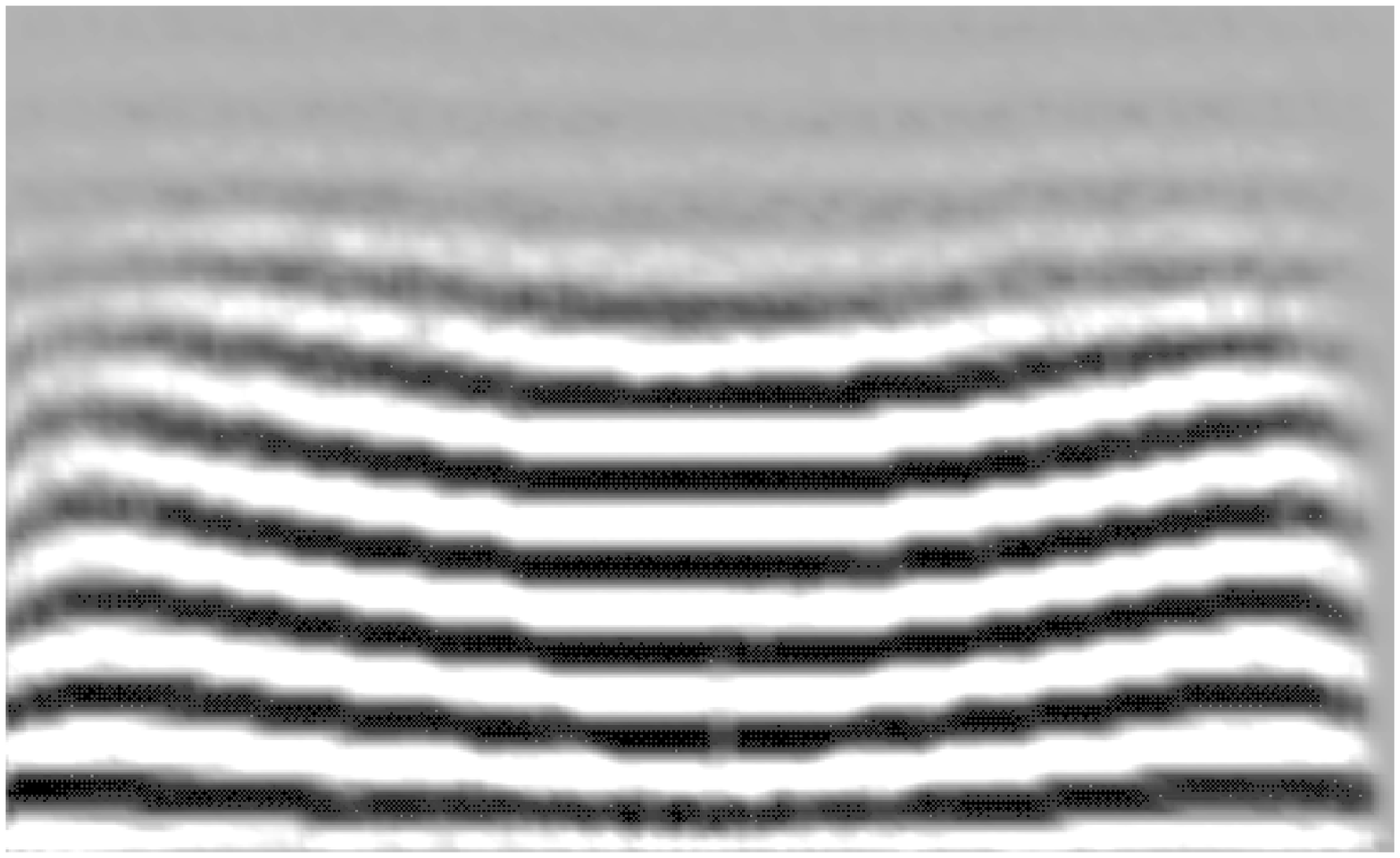,width=8cm,angle=0}} 
        \caption[capt6]
        { \label{fig:NMR} (above) Computer images simulating the motion
        of vertical (left) and horizontal (right) layers of granular
        material in the course of vertical shaking; (below) the same
        for horizontal shaking. }        
        \end{figure}

Consider the motion of a test granular particle in any averaged velocity
field generated by the procedure exposed above (different container shapes, 
various types of shaking). This is of practical interest in order to determine,
for example,
the effectiveness of a mixing device. In the absence of 
granular temperature, the motion of the test particle
would follow the flux lines. Starting from a given position, ${\bf x}_n$, 
after one period a new location ${\bf x}_{n+1}$ will be reached, which is
the result of the integration of elementary displacements (here equating
the number of frames per cycle). Fig. \ref{fig:NMR} shows the results
for vertical and horizontal shaking in the same  way as the Magnetic Resonance
Imaging\cite{NMR}. In this case computer imaging simulates the horizontal and 
vertical motion of colored bands initially parallel, which become distorted 
in the course of one period. 
The cycle-averaged motion is then reducible to iterative maps, leading to
mixing. Such mixing is incomplete  
since there may exist regions undergoing periodic motion, which
are not mixed at all. It happens, for example, in the center of convection
rolls accompanying horizontal shaking, potentially making the
horizontal shaking of thick layers ineffective for mixing particles.

The second origin of mixing is due to noise, leading to diffussion
which superimposes to the iterative map discussed above. The  
distance to diffuse in order to achieve complete
mixing is determined by the largest region containing limiting
cycles and/or fixed points.
Non-local noise at a length scale $l_n$ and time scale $t_n$ 
leads to nested displacements correlated in
space and time and helps to mix particles at distances exceeding
$l_n$ and times exceeding $t_n$. This subject has to be 
postponed until the properties of non-local noise are specified.
%-------------
%-------------
%%%%%%%%%%%%%%%%%%%%%%%%%%%%%%%%%%%%%%%%%%%%%%%%%%%%%%%%%%%%%%%%%%%%%%
  \section{CONCLUSIONS}

\begin{enumerate}

 \item Hydrodynamic equations provide an adequate theoretical frame for 
the study of dense granular systems. 

 \item Non-local or configurational noise adds on the ordinary hydrodynamic
noise, leading to non-selfaveraging properties in the limit of dense flows.

 \item Temperature-based mechanisms can be practically dismissed in the
description of dense gra\-nu\-lar flows.

 \item Averaging of data from molecular dynamic simulations is an useful
tool to reveal the details of the evolution of hydrodynamic fields. 

 \item Results of numerical integration of the system of hydrodynamic equations
show a qualitative agreement with the evolution of hydrodynamic fields.

 \item Study of fluctuations  and cycle-averaged motion concern practical 
applications such as mixing.
\end{enumerate}     
   
%%%%%%%%%%%%%%%%%%%%%%%%%%%%%%%%%%%%%%%%%%%%%%%%%%%%%%%%%%%%%
\acknowledgements     %>>>> equivalent to \section*{ACKNOWLEDGEMENTS}       
 
We wish to thank A. Malagoli, A. Patashinski and S. Nagel for useful 
conversations, and H. Jaeger for providing the file with the experimental
data published in Ref.~\citenum{Knight95}. S.E.E. is grateful
to T. Shinbrot for discussing practical aspects of granular mixing. 
C. S. acknowledges support by 
the Spanish Ministerio de Educacion y Cultura, grant number EX95-35065343.
At the initial stage the work of S.E.E. and T.P.
was supported by MRSEC Program of the
NSF under the Grant Number DMR-9400379.
%%%%%%%%%%%%%%%%%%%%%%%%%%%%%%%%%%%%%%%%%%%%%%%%%%%%%%%%%%%%%
%%%%% References %%%%%

%  \bibliography{spietext}   %>>>> bibliography data in spietext.bib

\begin{thebibliography}{1}

\bibitem{Haff83}
P.~K. Haff, ``Grain flow as a fluid mechanical problem,'' {\em J.~Fluid Mech.}
  {\bf 134}, pp.~401--430, 1983.

\bibitem{Haff86}
P.~K. Haff, ``A physical picture of kinetic granular flows,'' {\em J.~Rheology}
  {\bf 30}, pp.~931--948, 1986.

\bibitem{Vogel-Fultcher}
N.~F. Mott and E.~A. Davis, {\em Electronic Properties of Non-Crystalline
  Materials}, Oxford U. Press, New York, 1979.

\bibitem{Knight95}
J.~B. Knight, C.~G. Fandrich, C.~N. Lau, H.~M. Jaeger, and S.~R. Nagel,
  ``Density relaxation in a vibrated granular material,'' {\em Phys. Rev. E}
  {\bf 51}, pp.~3957--3963, 1995.

\bibitem{NMR}
E.~E. Ehrichs, H.~M. Jaeger, G.~S. Karczmar, V.~Y. Kuperman, and S.~R. Nagel,
  ``Granular convection observed by magnetic resonance imaging,'' {\em Science}
  {\bf 267}, pp.~1632--1634, 1995.

\bibitem{ESP96}
S.~E. Esipov, C.~Saluena and T.~Poschel, ``Granular glasses and fluctuational
  hydrodynamics.'' Preprint.

\end{thebibliography}
%  \bibliographystyle{spiebib}   %>>>> makes bibtex use spiebib.bst

  \end{document}